\documentclass[a4paper,11pt]{scrartcl}
\usepackage{noname}
\usepackage{tagging}
\usetag{ILD}
\usepackage{subcaption}
\usepackage{hyperref}
\usepackage{wasysym}
\usepackage{slashed}
\usepackage{wrapfig,rotating}
\usepackage{pgfpages}
\usepackage{fancybox,graphicx}
\usepackage{graphbox}
\usepackage{epstopdf}
\AppendGraphicsExtensions{.eps.gz}
\title{
  Search for new particles at the ILC

}
\def\MyNoteNumber{}
\localrep{DESY-22-179\\}
\date{\today}

\addauthor{
  Mikael Berggren

}
{\institute{1}}

\addinstitute{1}{
    Deutsches Elektronen-Synchrotron DESY,\\
  Notkestr. 85, 22607 Hamburg, Germany

}
\onbehalfof{
    \tagged{POS}{On behalf of }the ICFA-IDT-WG3 BSM group

}

\abstract{
  The LHC experiments have searched for and excluded many proposed
Beyond the Standard Model (BSM) theories. However, there are many scenarios
where LHC has little or no sensitivity.

Electron-positron colliders offers a different avenue for searches for such
phenomena.
In this talk, we will review the expectations for searches
for number of BSM models
at
the International Linear Collider (ILC).
We will discuss   new Higgs-like scalars, indirect
BSM via
Standard Model Effective Field Theory (SMEFT) fits, mono-photons,
and
SUSY in strongly or moderately compressed
models.

}

\titlecomment{%
  Contribution to the
   41st International Conference on High Energy physics - ICHEP2022\\
  6-13 July, 2022\\
  Bologna, Italy

}

\def\leqsim{\mathbin{\;\raise1pt\hbox{$<$}\kern-8pt\lower3pt\hbox{$\sim$}\;}}
\def\geqsim{\mathbin{\;\raise1pt\hbox{$>$}\kern-8pt\lower3pt\hbox{$\sim$}\;}}

\def\XPM#1{\mbox{$ \tilde{\chi}^{\pm}_#1                                $}}

\def\XN#1{\mbox{$ \tilde{\chi}^0_#1                                     $}}

\def\p#1{\mbox{$ \mbox{\bf p}_1                                         $}}

\newcommand{\smu}     {\mbox{$ \tilde{\mu}                                 $}}

\newcommand{\sel}     {\mbox{$ \tilde{\mathrm e}                           $}}

\newcommand{\stau}    {\mbox{$ \tilde{\tau}                                $}}

\newcommand{\eeto}    {\mbox{$ {\, \mathrm e}^+ {\mathrm e}^- \to             $}}

\newcommand{\GeV}     {\mbox{$ {\mathrm{GeV}}                              $}}

\newcommand{\TeV}     {\mbox{$ {\mathrm{TeV}}                              $}}

\newcommand{\ba}{\begin{array}}
\newcommand{\ea}{\end{array}}
\newcommand{\bc}{\begin{center}}
\newcommand{\ec}{\end{center}}
\newcommand{\be}{\begin{eqnarray}}
\newcommand{\eeq}{\end{eqnarray}}
\newcommand{\bes}{\begin{eqnarray*}}
\newcommand{\ees}{\end{eqnarray*}}
\newcommand{\Kz}{\ifmmode {\rm K^0_s} \else ${\rm K^0_s} $ \fi}
\newcommand{\Zz}{\ifmmode {\rm Z^0} \else ${\rm Z^0 } $ \fi}
\newcommand{\xxbar}{\ifmmode {\rm x\bar{x}} \else ${\rm x\bar{x}} $ \fi}
\newcommand{\rphi}{\ifmmode {\rm R\phi} \else ${\rm R\phi} $ \fi}

\def    \missEt      {\ifmmode{/\mkern-11mu E_t}\else{${/\mkern-11mu E_t}$}\fi}
\def    \missE       {\ifmmode{/\mkern-11mu E}\else{${/\mkern-11mu E}$}\fi}
\def    \missp       {\ifmmode{/\mkern-11mu p}\else{${/\mkern-11mu p}$}\fi}
\def    \misspt      {\ifmmode{/\mkern-11mu p_t}\else{${/\mkern-11mu p_t}$}\fi}

\begin{document}
\titlepage

\section{The strong points of ILC for searches}

\begin{figure}[h] 
  \begin{center}
  \includegraphics [scale=0.3]{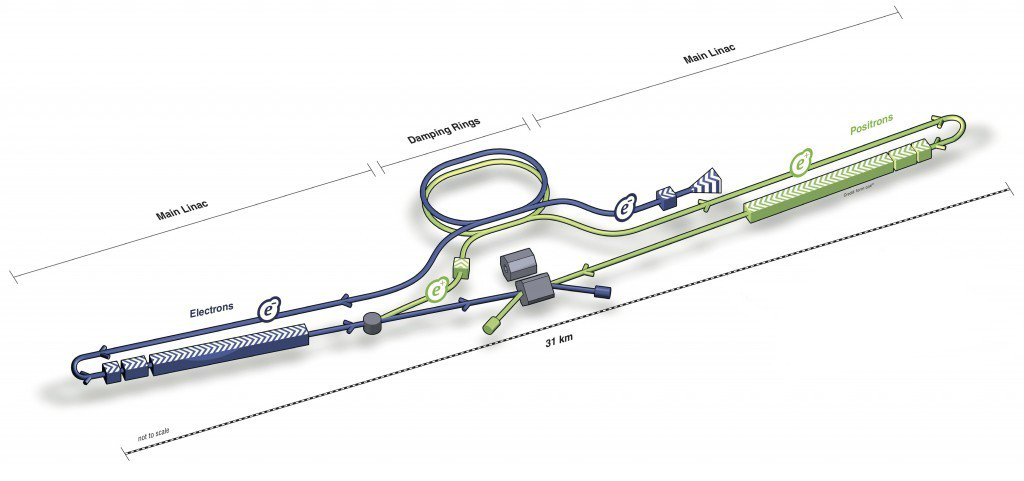}
  \includegraphics [scale=0.3]{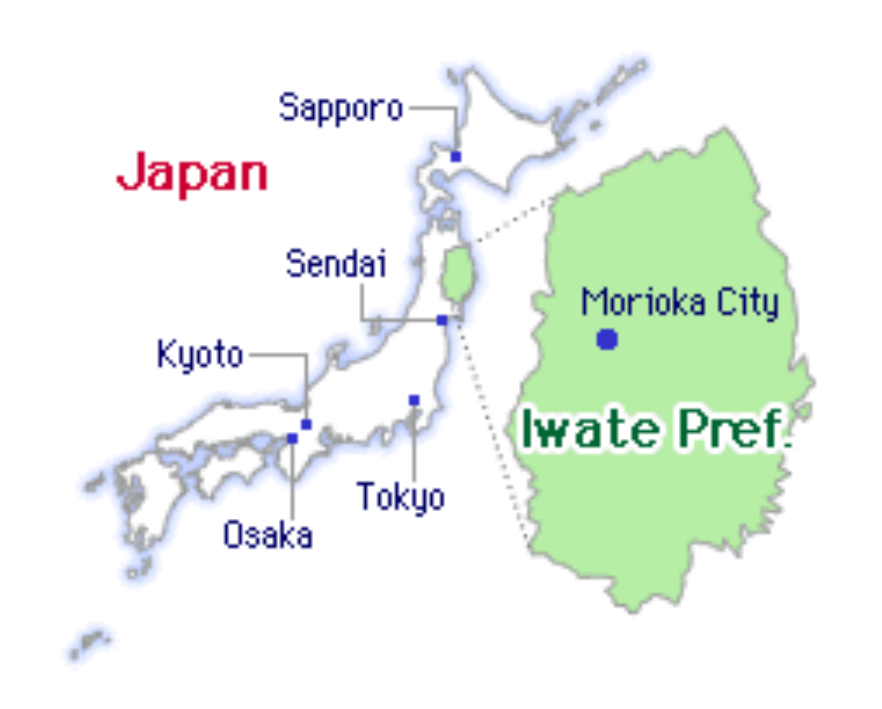}
   \end{center}
 \caption{Schematic of the ILC and the location of the proposed site in Japan's Tohoku region.\label{fig:ilc}}
 \end{figure}

The proposed International Linear Collider (the ILC \cite{Adolphsen:2013kya}, Fig.~\ref{fig:ilc}) would
collide polarised electrons with polarised positrons.
The centre of mass energy, E$_{CMS}$, will initially be  250 \GeV~, and then extended to 500 \GeV. Possibilities
to go to 1 \TeV,
and operate at $E_{CMS}= M_Z$ also exist.
That the  initial state is e$^+$e$^-$ implies electroweak production, which leads to low background rates.
This is an advantage for the detector design and optimisation: Since the detectors would not need to be radiation-hard,
the tracking system can be realised with a total thickness as low as a few percent of
a radiation-length.
The low rates means that the detectors do not need to be triggered, so that \textit{all}
events produced  will be recorded.
In addition, the detector system can have a coverage of nearly $ 4\pi$.
Since point-like objects are colliding,  the initial state is fully  known at an   e$^+$e$^-$ machine.
This will be quite important for many searches for new phenomena.
The  ILC has a  20 year running plan defined, with 
programmes giving integrated luminosities of 2 and 4 ab$^{-1}$  at $E_{CMS}=$ 250 and 500 \GeV, respectively.
It could provide  8  ab$^{-1}$ at the possible upgrade to 1 \TeV.  
The construction of the ILC is currently subject to a high-level political consideration in Japan.

 To perform searches or measurements Beyond the Standard Model (BSM) as well as precision
 measurements of the standard model (SM)
 requires that the excellent conditions offered by the machine are matched by equally out-standing detector capabilities.
 Specifically,
 a jet energy resolution of  3-4\%,
 an asymptotic  momentum resolution of $\sigma(1/p_\perp) = 2 \times 10^{-5}$ \GeV $^{-1}$,
 and a measurement of impact-parameters better than 5 $\mu \mathrm{m}$ will be required \cite{ILC:2007vrf}.
 The detectors should be hermetic,
 with no other gaps in the acceptance than the
 unavoidable beam-pipes bringing the electron and positrons into them.
 In addition, the detectors must be able to record data without being triggered,
 and to avoid the need for active cooling, they must be able to operate
 with the electronics switched off between  bunch-trains.
 The two detector concepts currently under consideration for the ILC -
 the International Large Detector concept (the ILD)\cite{ILDConceptGroup:2020sfq} and
 the Silicon Detector (SiD) \cite{Behnke:2013lya} - are equipped with
 high granularity calorimeters
 optimised for particle flow \cite{Thomson:2009rp} allowing the required resolution of the jet energy.
 Extensive simulations show that both concepts can meet all requirements \cite{ILDConceptGroup:2020sfq,Behnke:2013lya}.
  
    
\section{BSM at ILC: New scalars, small deviations from the SM, mono-photons}

\begin{figure}[t]
  \begin{center}
    \subcaptionbox{}{\includegraphics [align=c,scale=0.26]{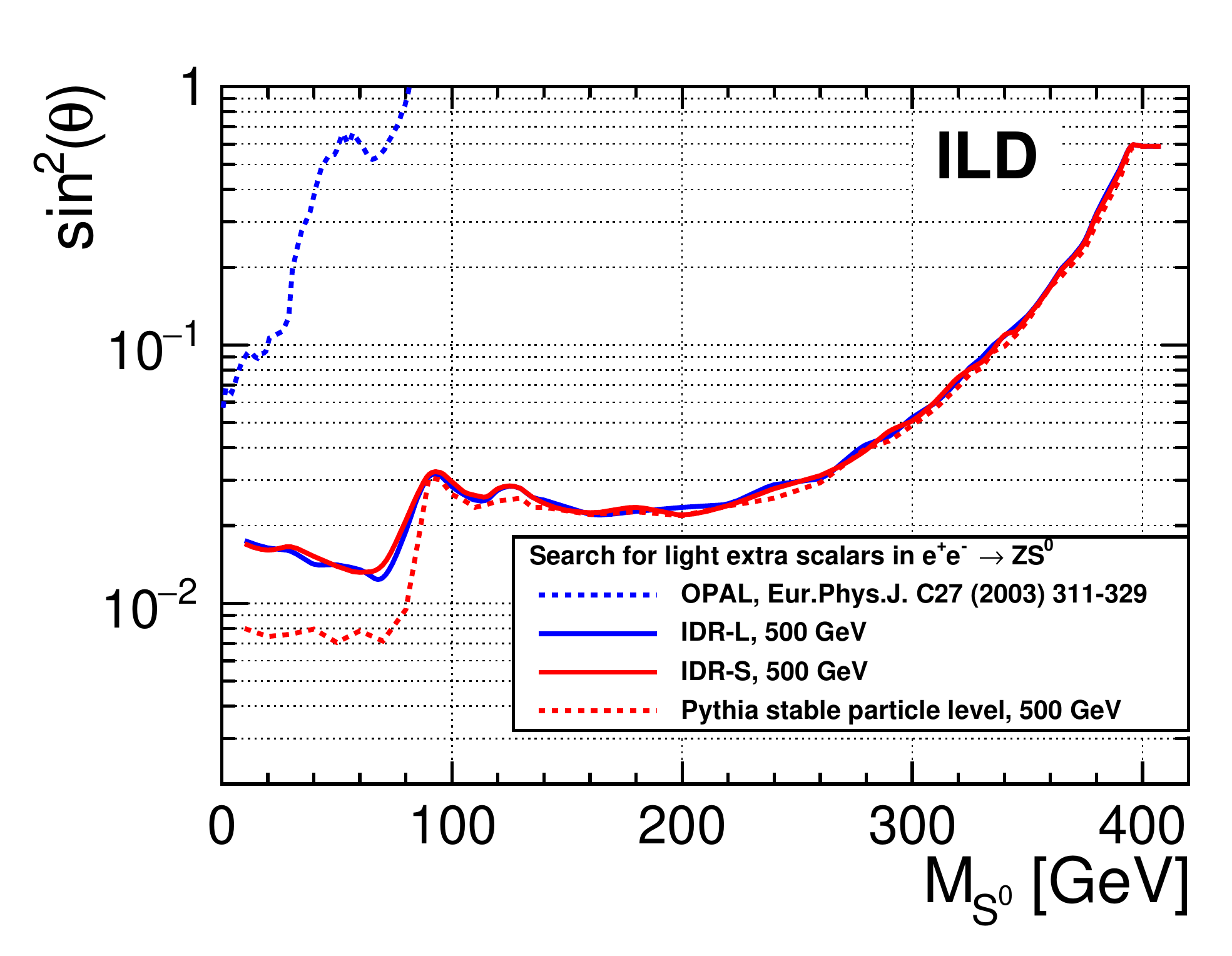}}
    \subcaptionbox{}{\includegraphics [align=c,scale=0.28]{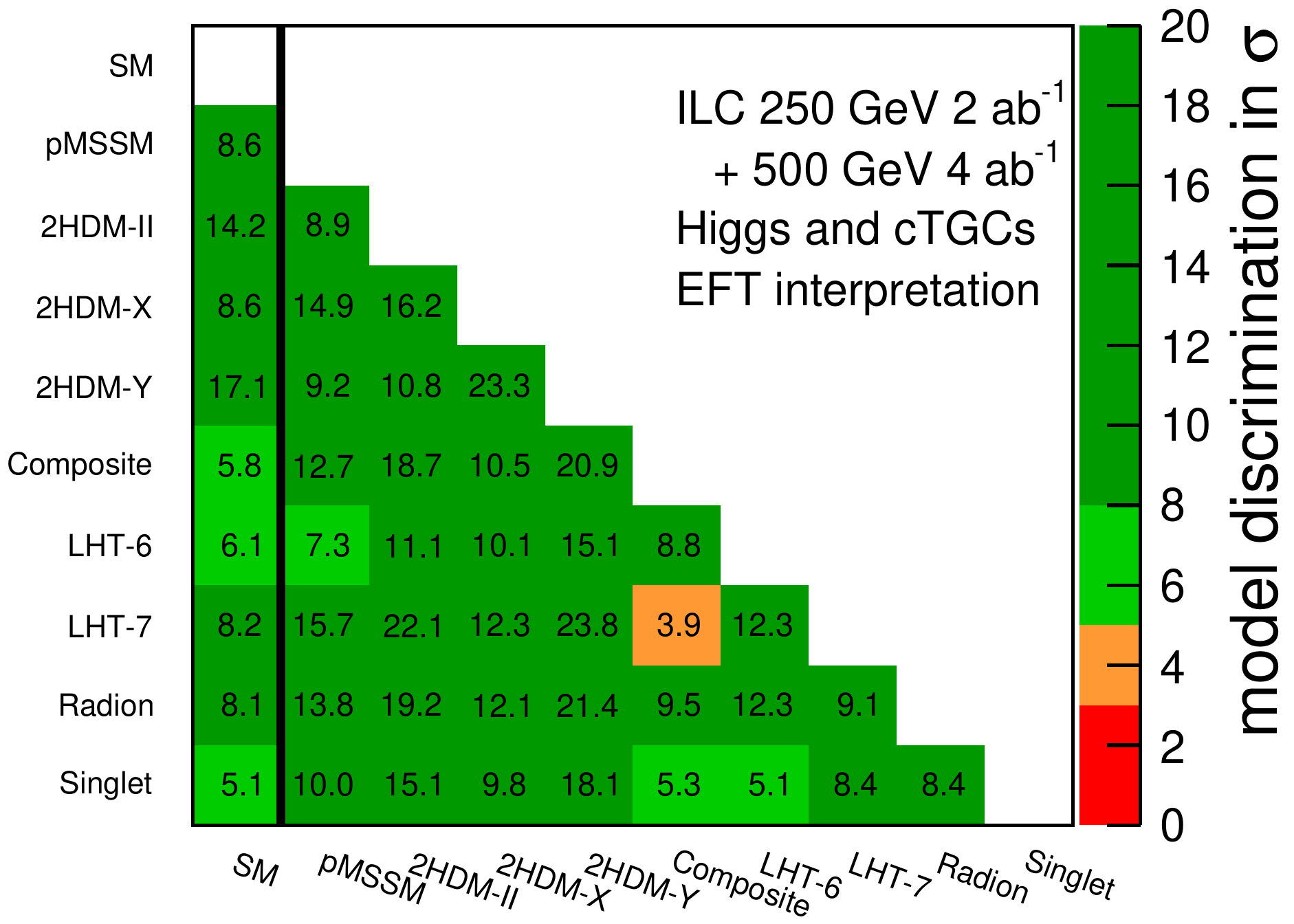}}
\end{center}
\caption{
  (a) Projected exclusion limit for
  new scalars, in terms of the coupling compared to the coupling an SM Higgs at the same mass would have.
  (b) Significances of SMEFT deviations from the expectation, both for the SM expectation and the expectation of
  each of the various listed models.
  \label{fig:otherbsm}}
 \end{figure}

Many BSM models predict the existence of a new Higgs-like scalar ($S$), produced in $\eeto Z^* $ $\rightarrow Z S$.
The decay modes of $S$ would not be known a priori.
Such a state could have escaped detection at LEP if its production cross-section is much less than that of a SM Higgs of the same mass.
Therefore, a search for $S$ should be performed at all available masses, and without any assumptions on the decay modes.
At ILC, this search can be made using the recoil-mass, i.e. the mass of the  system recoiling against the measured $Z$.
In \cite{Wang:2020lkq}, a full
detector simulation study was performed, and it was found
that couplings down to a few percent of the  SM-Higgs equivalent can be excluded,  see Fig. \ref{fig:otherbsm}(a).

The ILC will also be powerful in detecting BSM through indirect searches,
i.e. by observing deviations from the  behaviour predicted by the SM.
These deviations can not only be detected, but they can also often be used to separate BSM models.
An example of this type of analysis of BSM physics is illustrated in Fig. \ref{fig:otherbsm}(b) where we show a
Standard Model Effective Field Theory (SMEFT) study  \cite{Barklow:2017suo} using ILC results on Higgs properties and
triple gauge couplings (TGCs).
The authors have selected models that cannot be detected at the HL-LHC.
One can note that ILC would be able to separate all the models from the SM (at the 5 $\sigma$ level),
and also to separate them from each other, at a similar  confidence level.
\begin{figure}[b]
\begin{center}
\subfloat[][]{
  \includegraphics[scale=0.28]{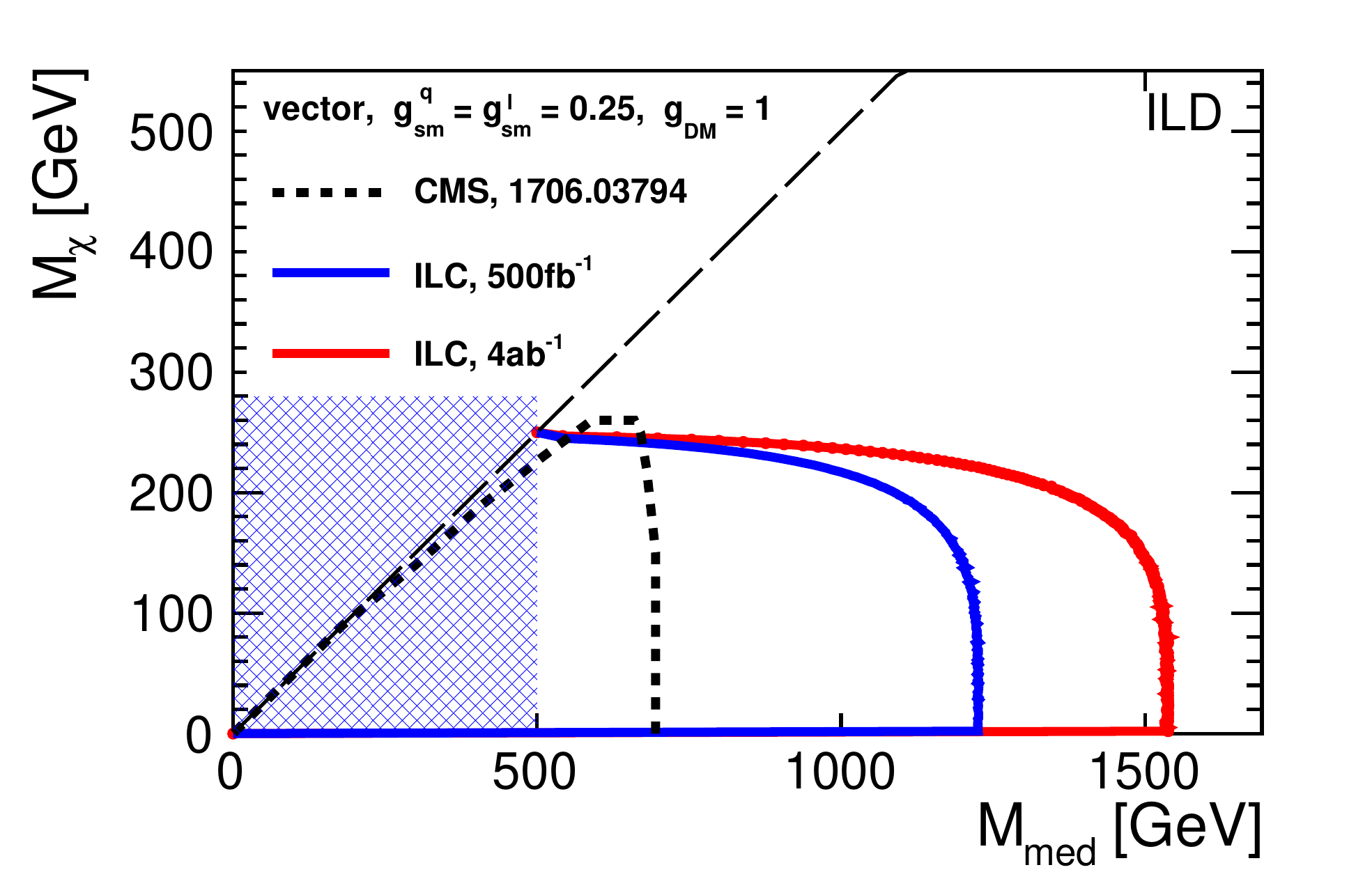}
  }
\subfloat[][]{
\includegraphics[scale=0.58]{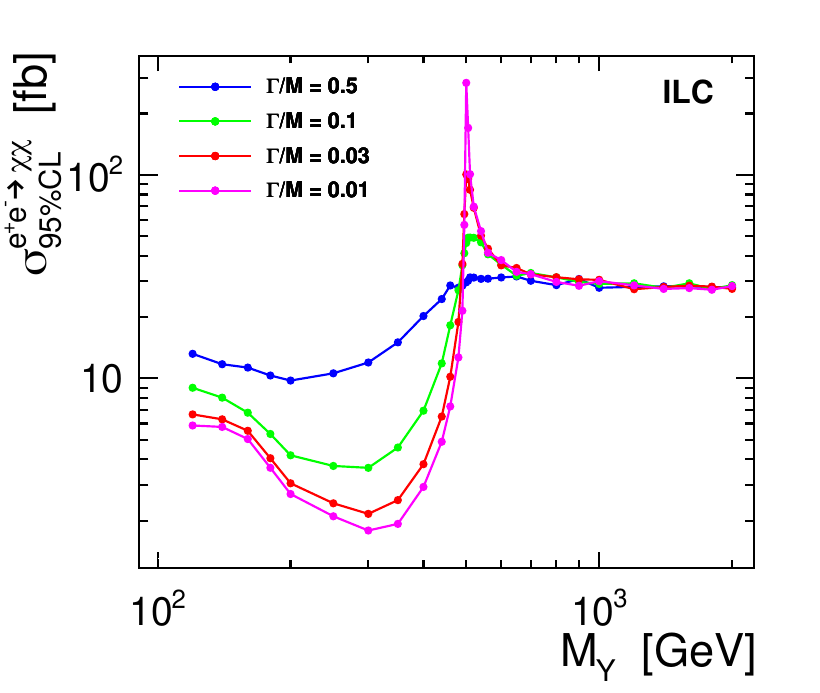}
}
\caption{Dark matter searches using the Mono-photon signature at ILC. (a) Heavy mediators, allowing for an EFT approach.
  (b) Arbitrary mediator masses, for various assumed widths of the mediator. \label{fig:monophot}}
\end{center}
\end{figure}

Dark Matter can be searched for at the ILC in the {mono-photon} channel, i.e.  in the process $\eeto (DM)(DM) +\mathrm{ISR} \gamma$.
Expected results are  shown in Fig. \ref{fig:monophot}(a) and (b), both for heavy mediators (a),
where a model independent  EFT approach is appropriate \cite{Habermehl:2020njb}, and for 
arbitrary mediators (b), where the sensitivity will depend on properties of the
mediator \cite{Kalinowski:2021tyr}. Both cases show potential beyond HL-LHC reach.

\section{BSM at ILC: SUSY}

  Supersymmetry (SUSY) \cite{Wess:1974tw} is the most complete theory of BSM, and should therefore
  be given special  attention.
    Naturalness, the hierarchy problem, the nature of dark matter (DM),
  or the observed value of the magnetic moment of the muon, are all reasons to prefer
  a light electroweak sector of SUSY.
  Moreover, many models
and the global set of constraints from observation
  point to a \textit{compressed spectrum}.
  If the Lightest SUSY Particle (the LSP) is a Higgsino or a Wino, there must be other
  bosinos  with a mass close to it, since the $\tilde{H}$ and $\tilde{W}$
  fields have several components, leading to a close connection between
  the physical states of the bosinos.
  Although the third possibility -  a Bino-LSP - has no such constraints,
  an overabundance of DM  is expected in this case \cite{Roy:2007ay}.
  To avoid such a situation,
  a balance between early universe LSP production and
    decay is needed.
  \begin{figure}[t]
    \begin{center}
     \subcaptionbox{}{\includegraphics [align=t,scale=0.28]{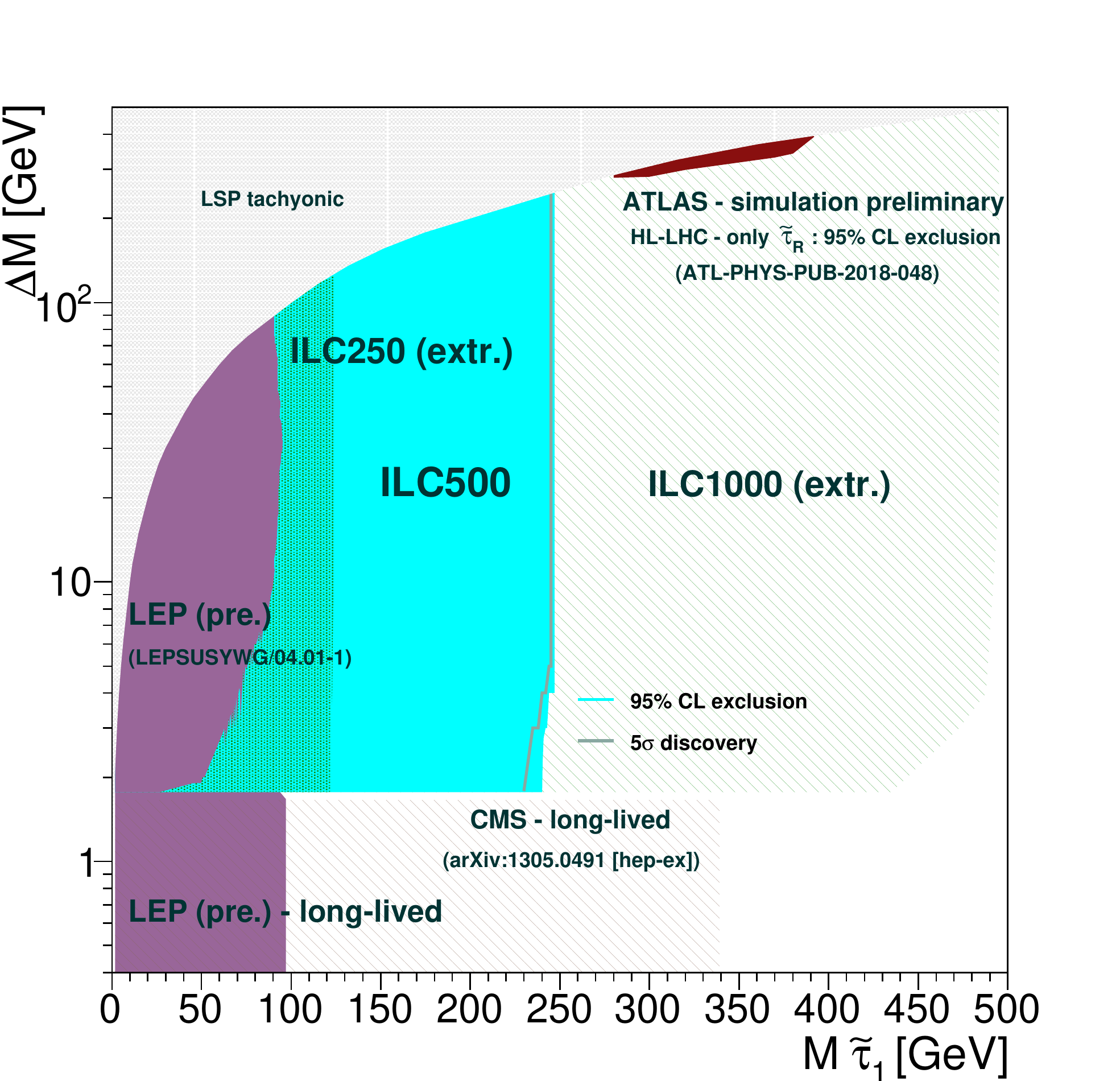}}
\subcaptionbox{}{\includegraphics [align=t,scale=0.30]{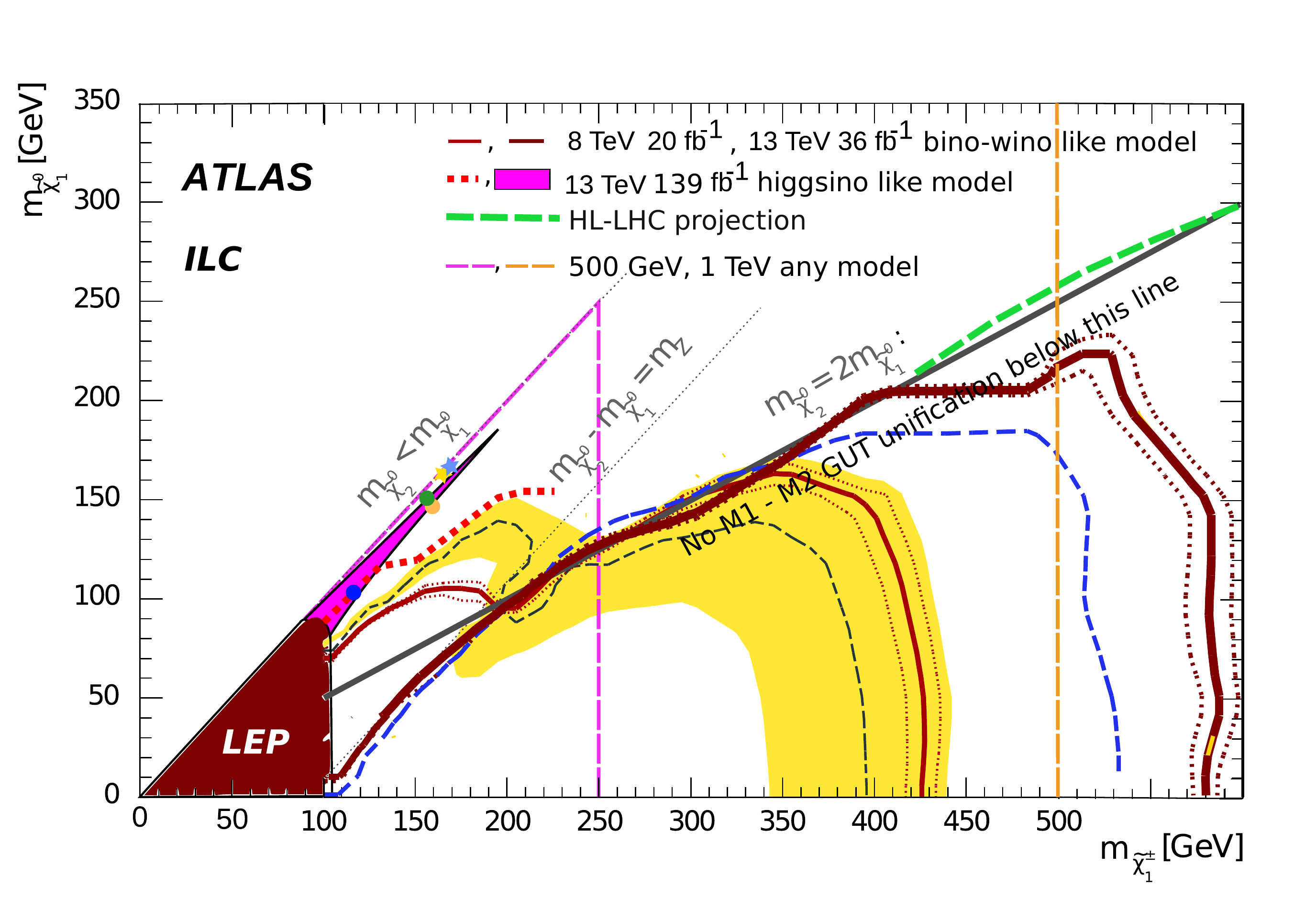}}
\end{center}
\caption{ Observed or projected exclusion regions for a $\stau$ (a) or a $\XPM{1}$ (b) NLSP, for LEPII, LHC, HL-LHC and for ILC-500 and ILC-1000. For the
  ILC curves, discovery and exclusion reach are both within the width of the lines.\label{fig:X1summary}
}
 \end{figure}
  One compelling option is $\stau$~co-annihilation \cite{Roy:2007ay}, and 
  for this process to contribute sufficiently, the density in the early universe of $\stau$ and $\XN{1}$
  must be close,
  which means that their masses must be quite similar.
  
Most sparticle-decays occur via cascades. In the case of compressed spectra, 
  the last decay in the cascade - the one to SM particles and the LSP -
  has a small $\Delta M$, and hence low visible activity.
  For such decays, the current limits from LHC  are for specific models,
  and only the LEPII limits are model-independent.
Indeed, current observations  from LHC run 2, LEP, g-2, DM (assumed to be 100\%~LSP),
and precision observables together also suggest
a compressed spectrum~\cite{Bagnaschi:2017tru}.
 
At the ILC, it is possible to perform a loophole free SUSY search,
since in SUSY, the properties of
the production and decay of NLSPs are fully predicted for given masses of the LSP and the NLSP.
All possible NLSP candidates can therefore be searched for in a systematic way.
In Fig. \ref{fig:X1summary} shows the current or projected limits 
for a $\stau$ NLSP (a) \cite{NunezPardodeVera:2021cdw}, or a $\XPM{1}$ one (b) \cite{PardodeVera:2020zlr,ATLAS:2018ojr}.
\begin{figure}[b] 
   \begin{center}
     \includegraphics [scale=0.21]{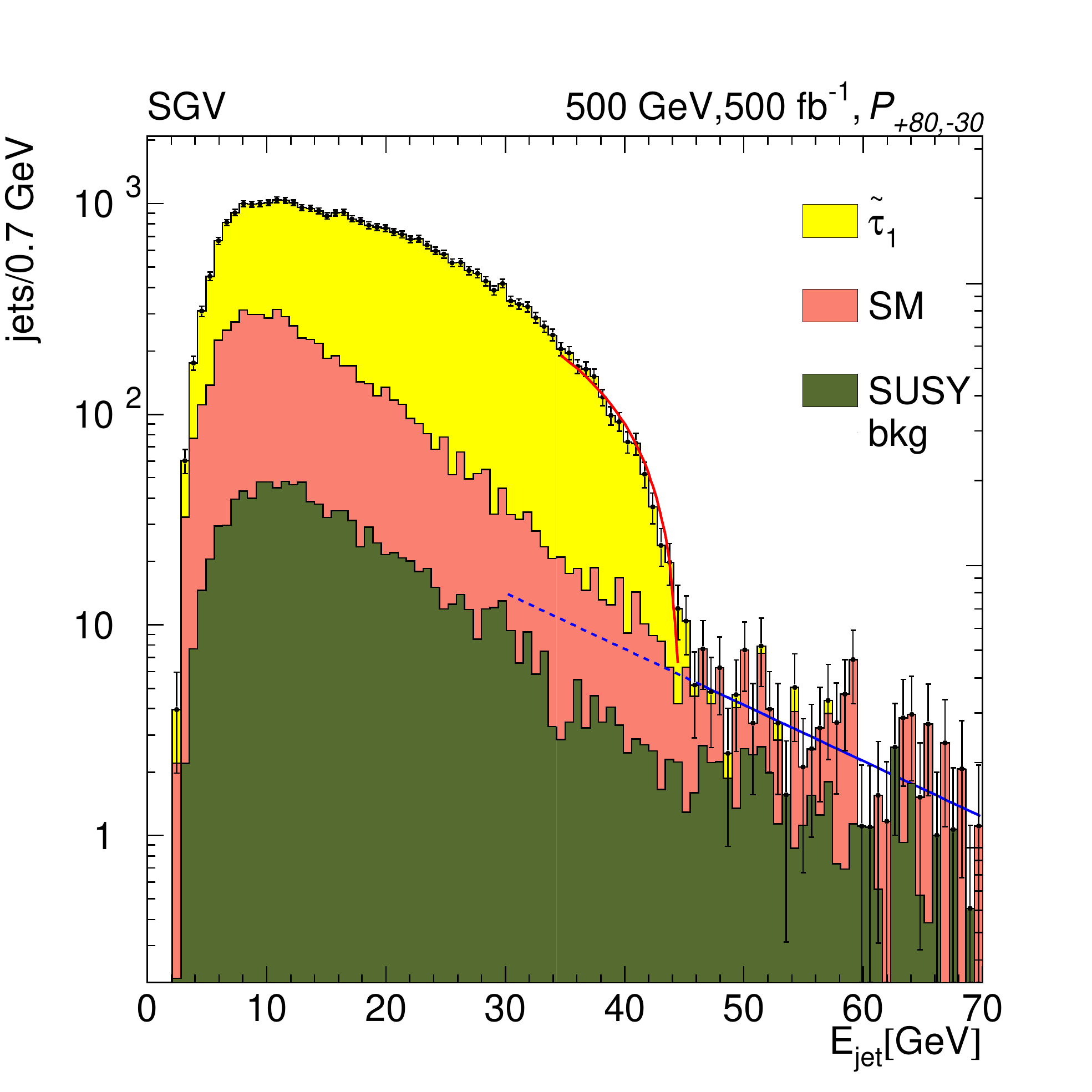}
     \includegraphics [scale=0.21]{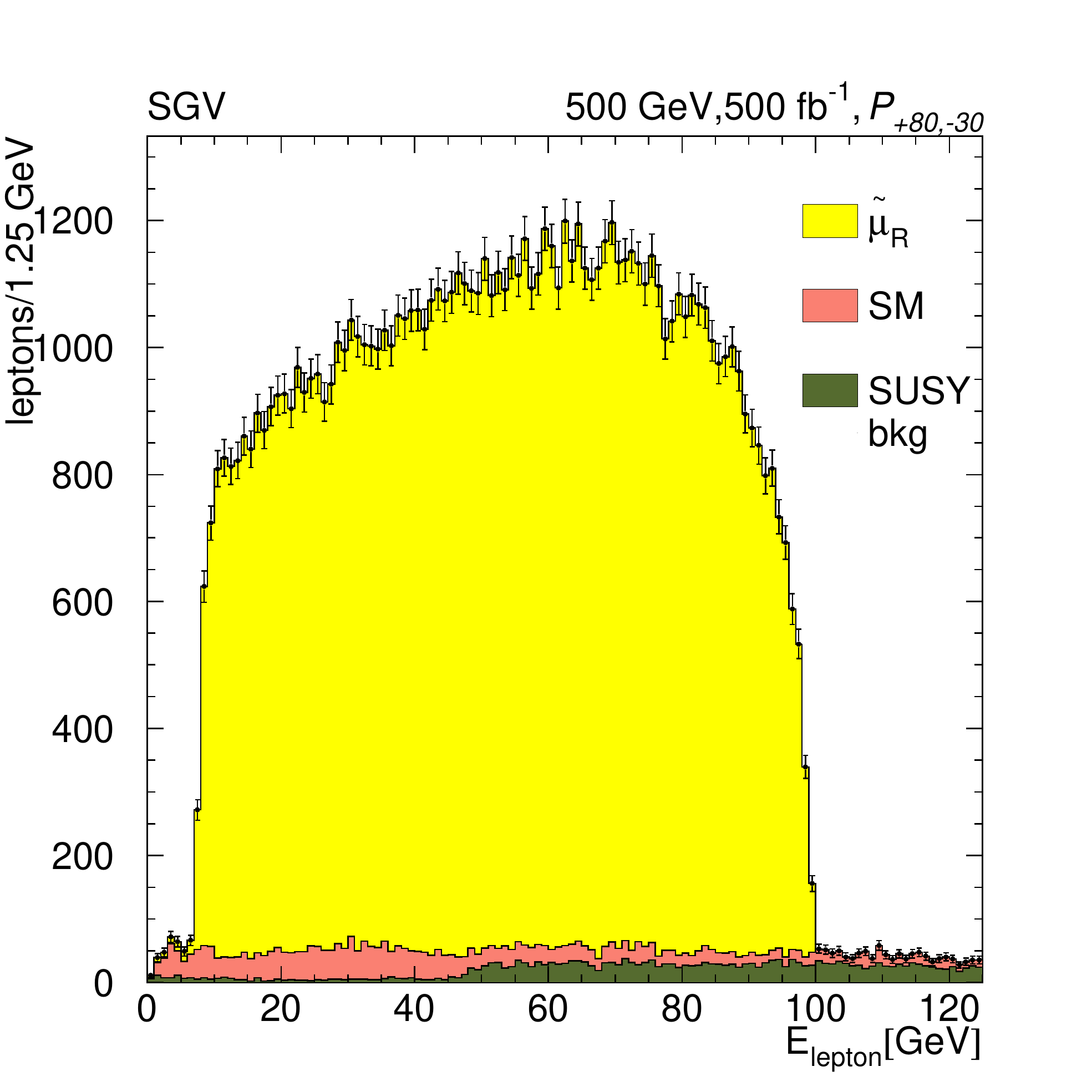}
     \includegraphics [scale=0.21]{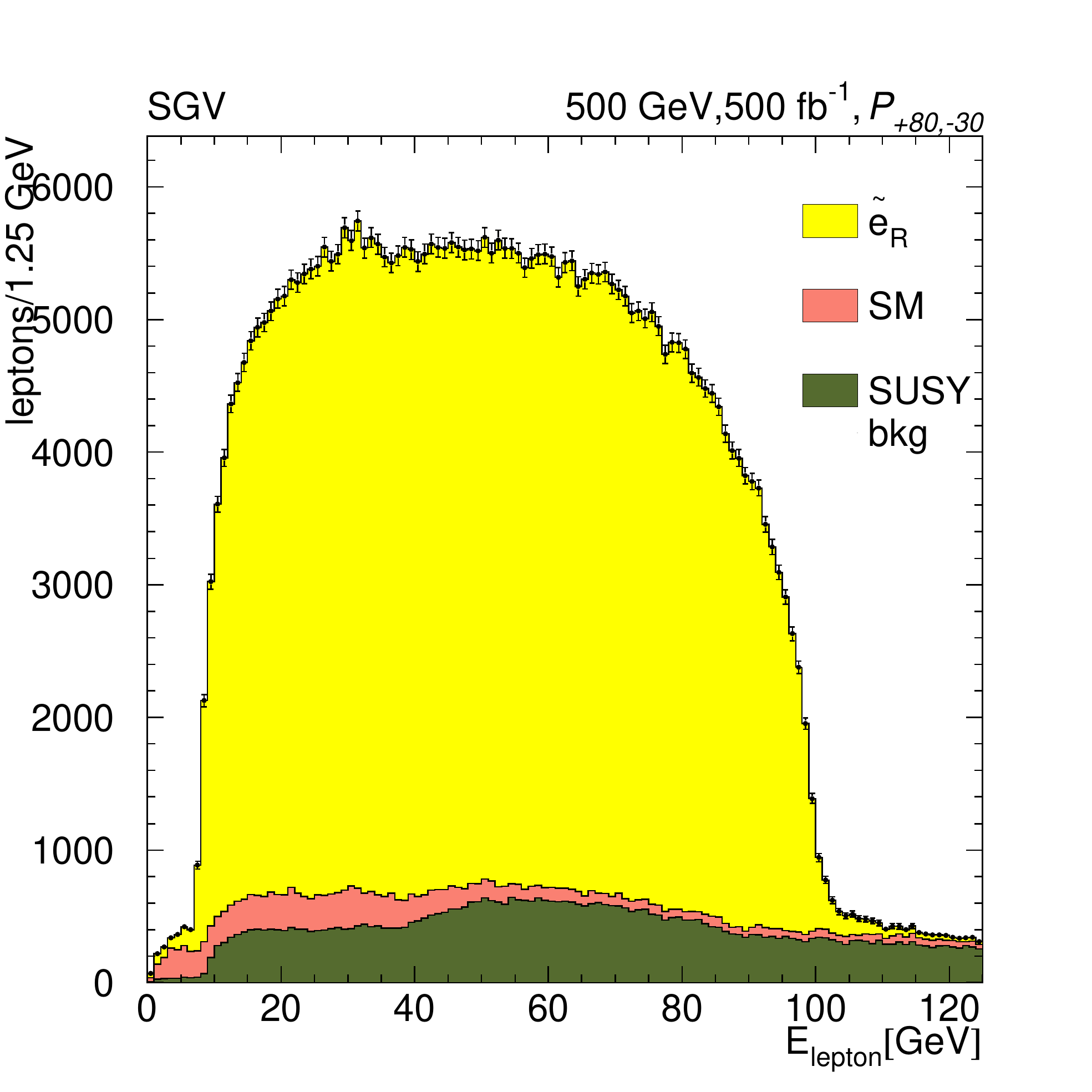}
     
      \includegraphics [align=c,scale=0.195]{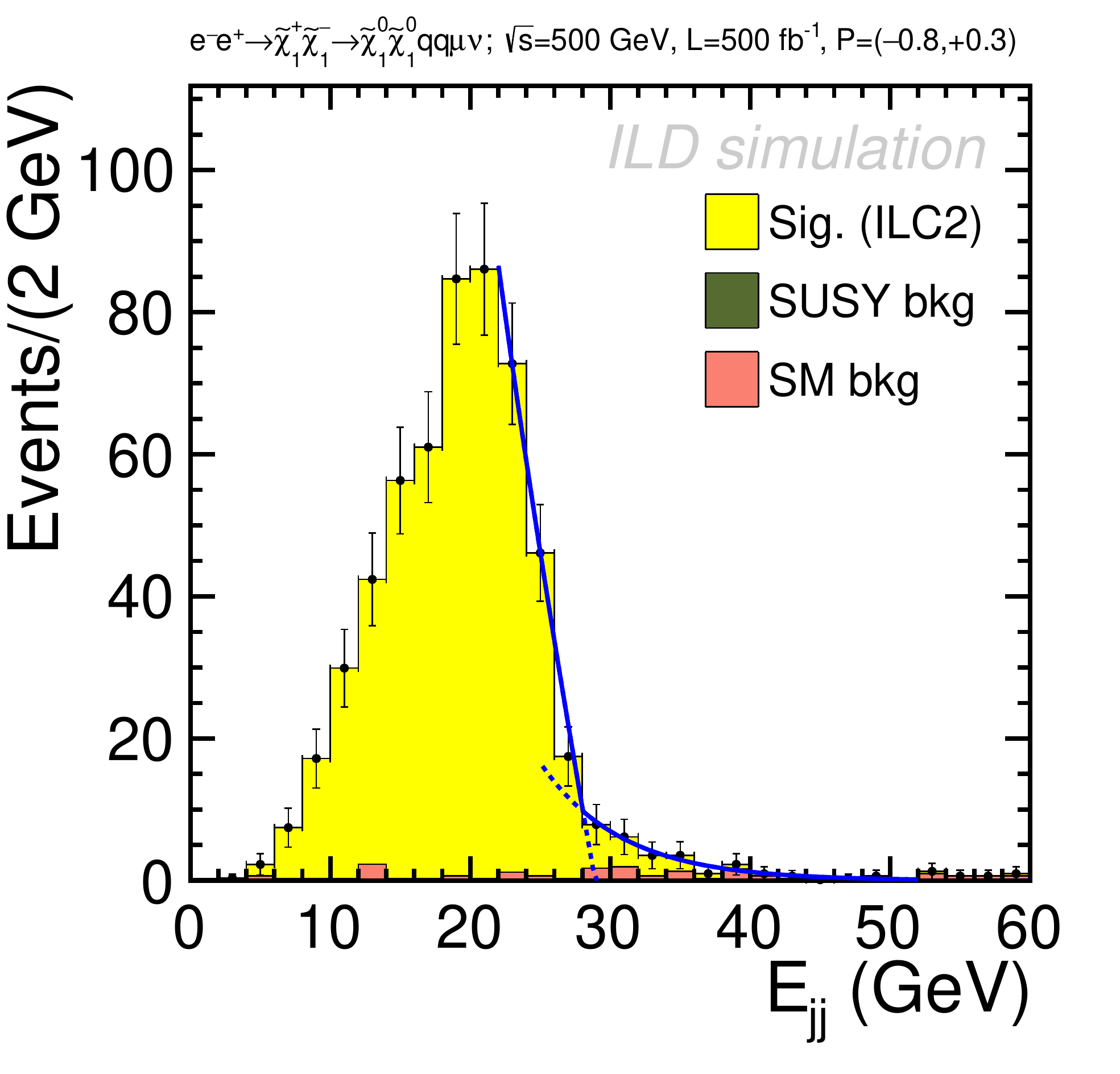}
      \includegraphics [align=c,scale=0.195]{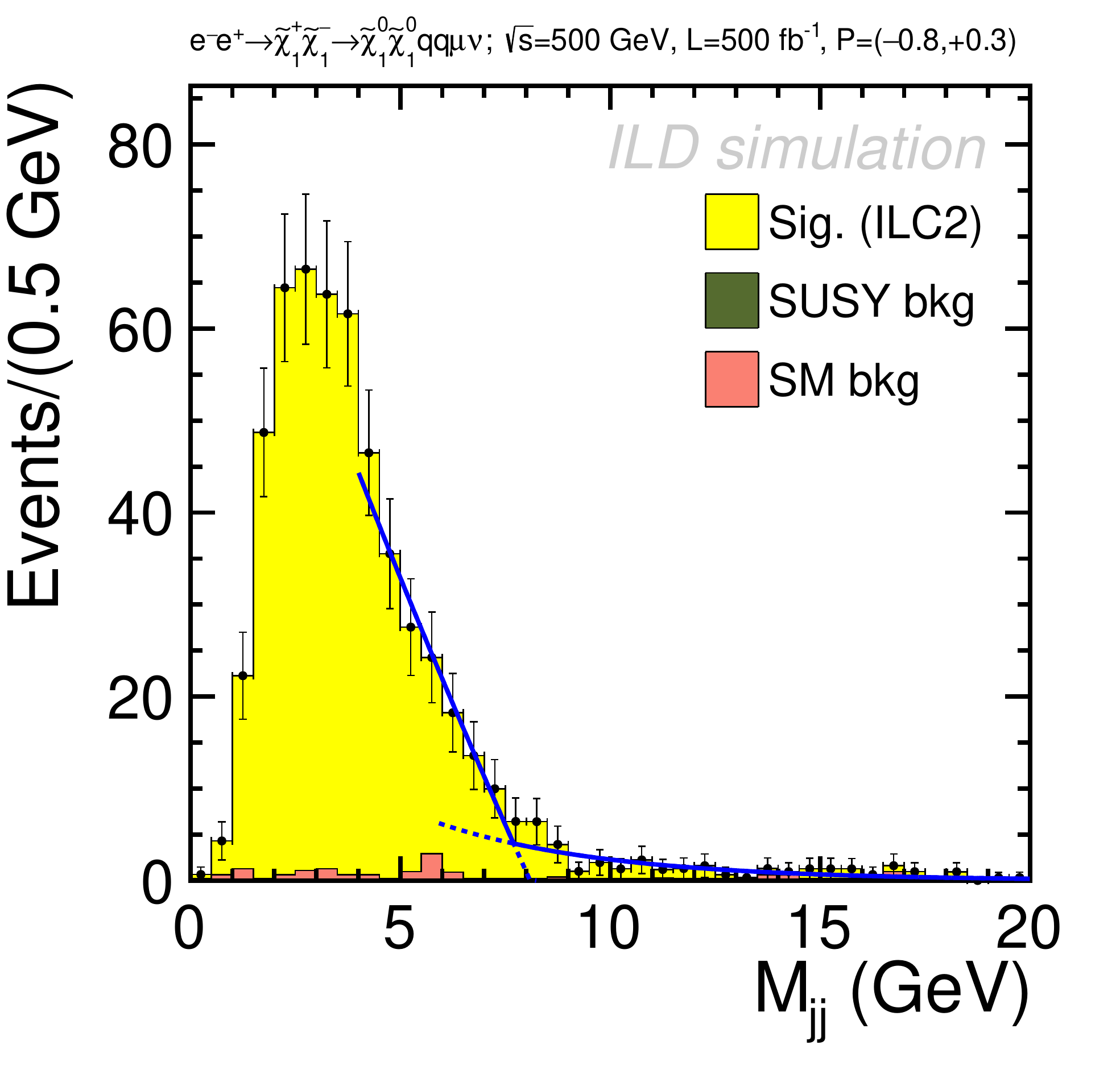}
      \includegraphics [align=c,scale=0.21]{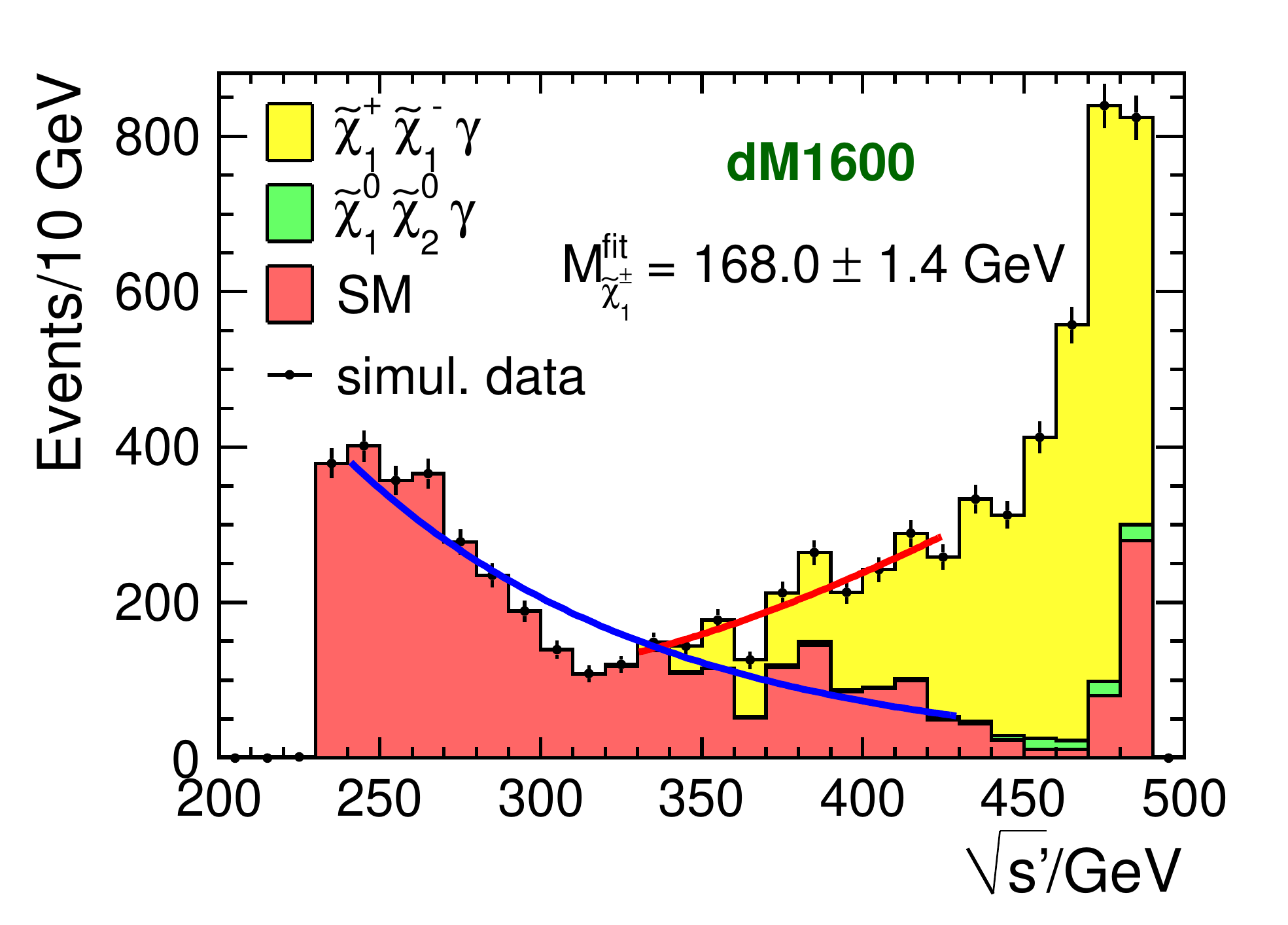}
      
      \includegraphics [align=c,scale=0.195]{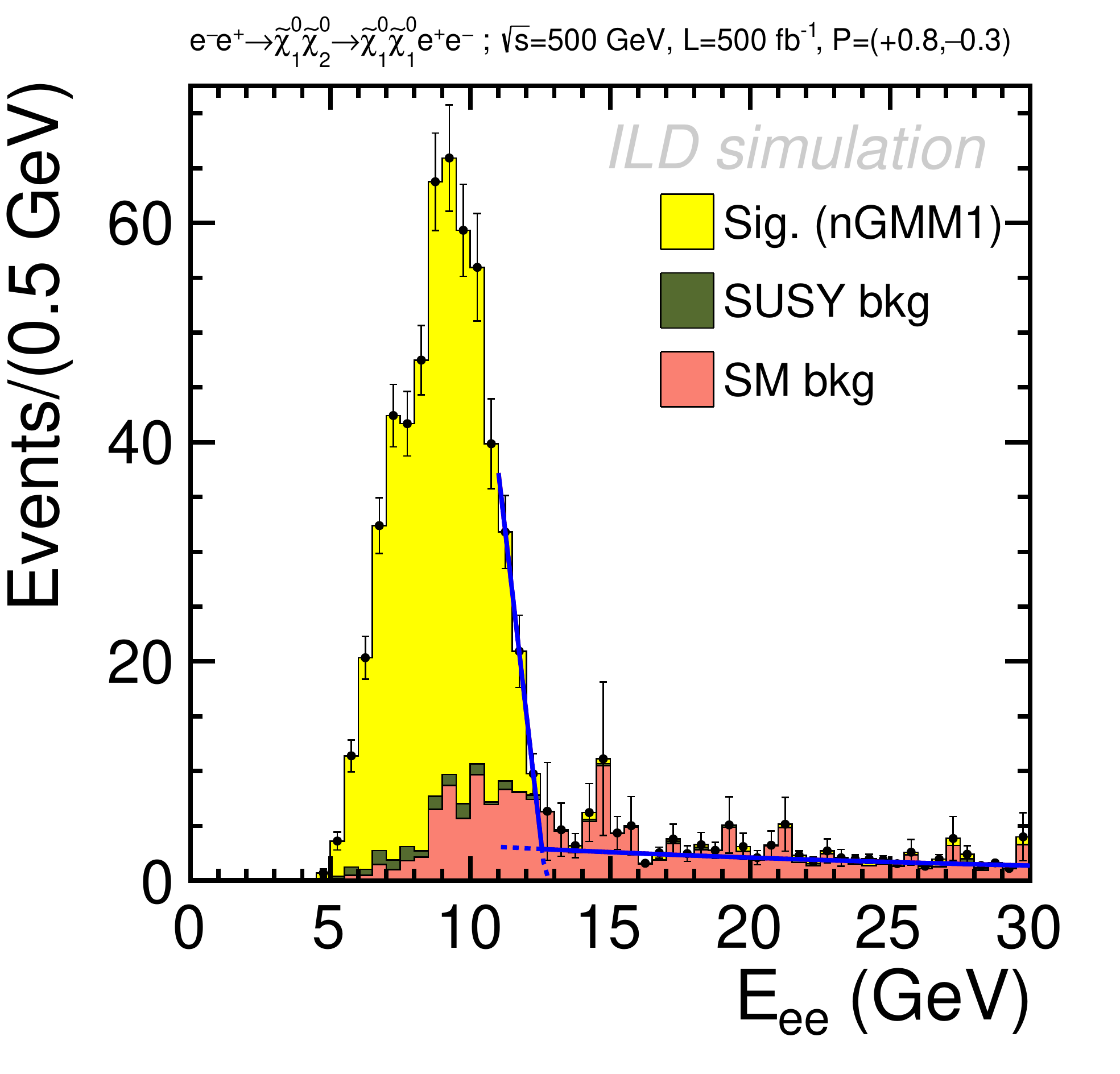}
      \includegraphics [align=c,scale=0.195]{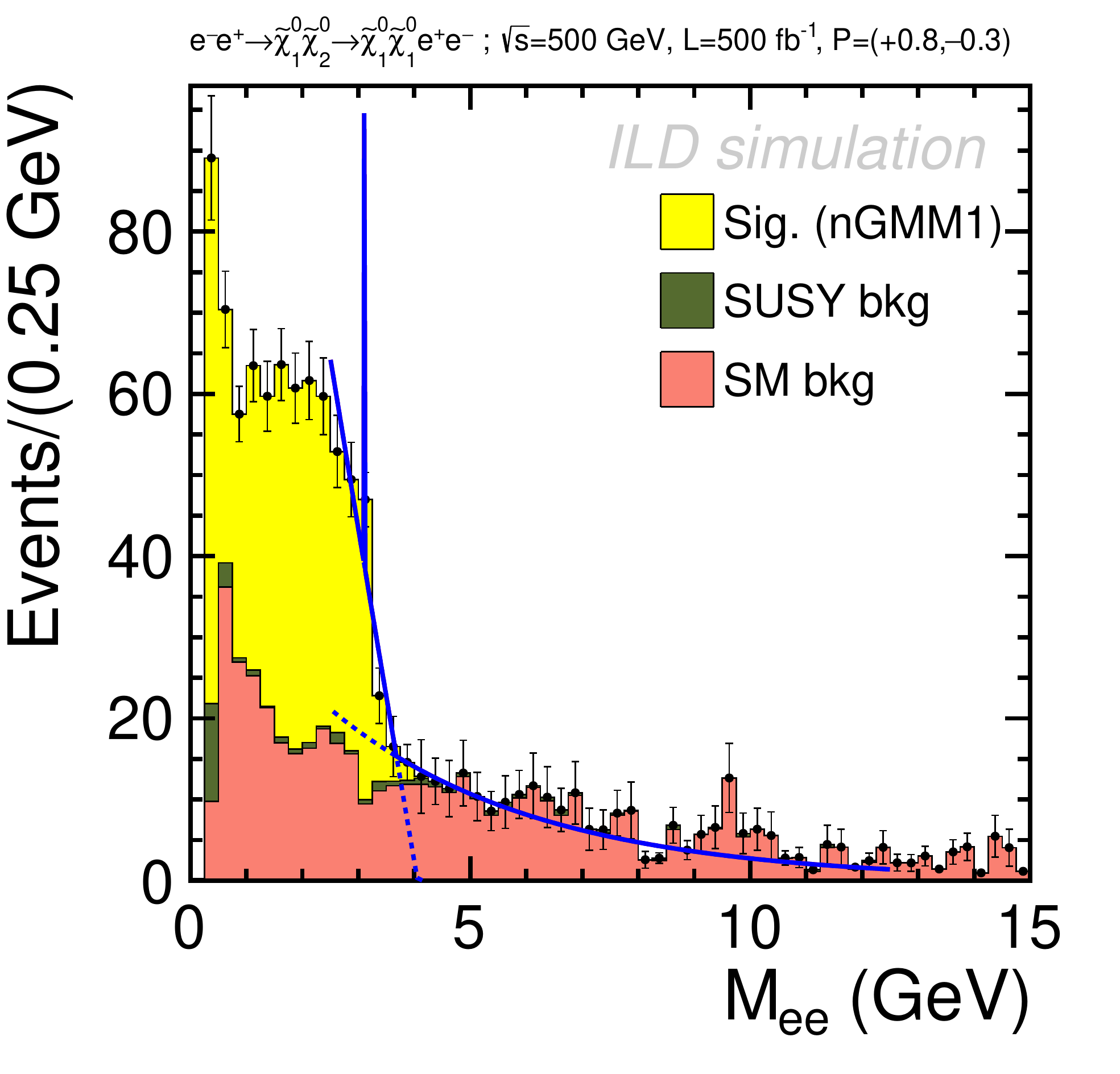}
      \includegraphics [align=c,scale=0.21]{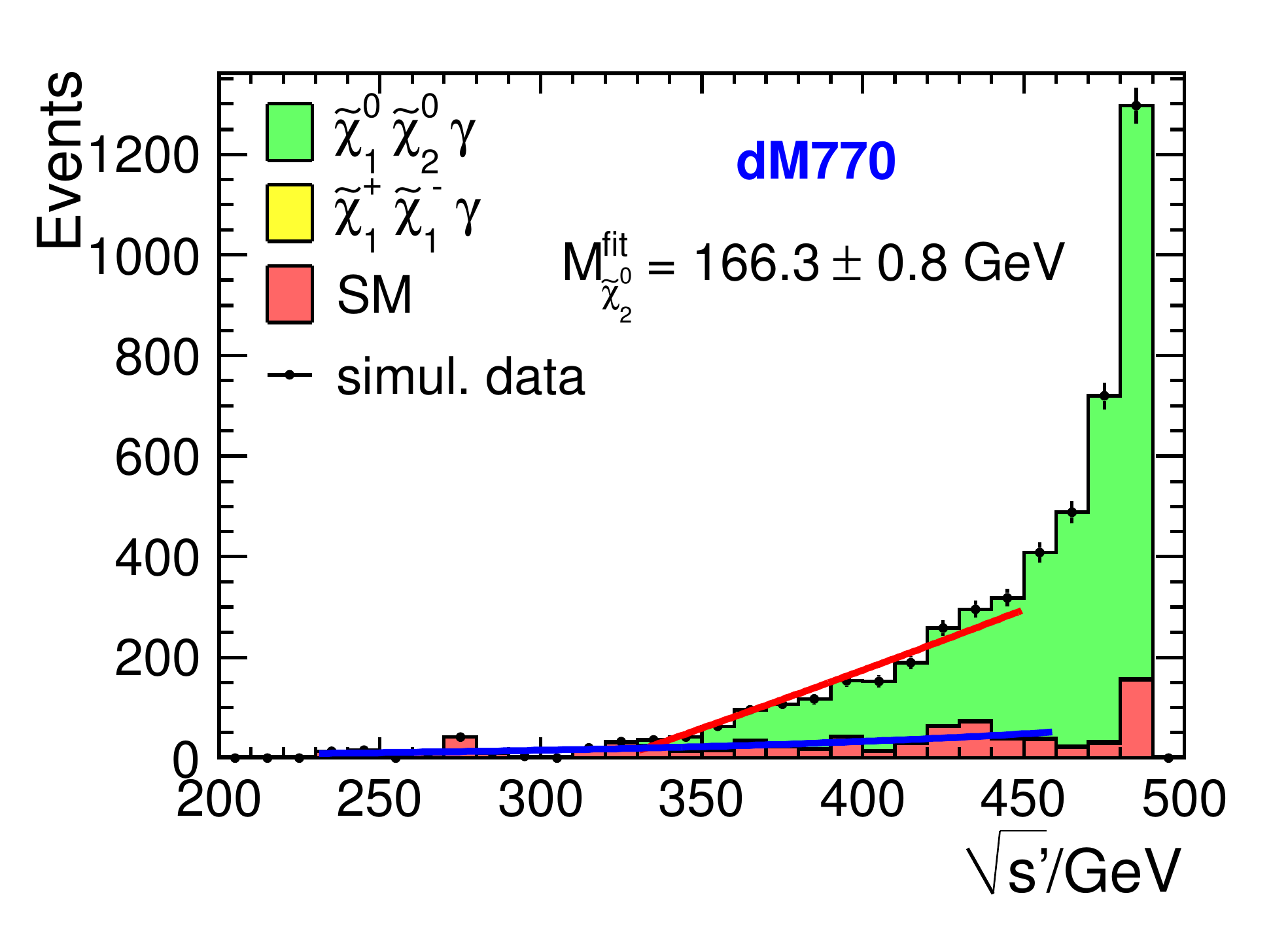}
    \end{center}
    \caption{Top row: $\stau$, $\smu$ and $\sel$ spectra. Middle and bottom rows: Observables
      for three different Higgsino-LSP models. The middle row shows the case of $\XPM{1}$ production, the bottom one 
      that of $\XN{2}$ production.  \label{fig:sleptC1N2}}
\end{figure}  
As can be seen in the figure, exclusion- and discovery-reach is very close at ILC,
so that, at the ILC, a SUSY discovery would take place quite quickly.
The situation that an interesting SUSY signal is at the intermediate
level (neither excluded, nor discovered) for years will never occur: Either the
process is not reachable and there is no sign of it, or it will be discovered immediately.
This means that SUSY studies at the ILC would almost immediately enter the
realm of precision studies. The plots in Fig. \ref{fig:sleptC1N2} shows a number of examples of the
type of signals that can be expected:
 Typical slepton signal ($\stau$, $\smu$ and $\sel$) in the top row,
       in a  $\stau$ co-annihilation model (FastSim) \cite{Berggren:2015qua}.
       Typical chargino and neutralino signals in different Higgsino LSP models are shown in the following rows.
       The two plots on the left are models  with moderate (a few to some tens GeV) $\Delta{M}$ (FullSim) \cite{Baer:2019gvu},
       while those on the  right are  for a  model with very low (sub-GeV) $\Delta{M}$ (Fast/FullSim) \cite{Berggren:2013vfa}.
      In all the cases illustrated,
 SUSY masses could be determined at the sub-percent level,
 the polarised production cross-sections at a level of a few percent.
 Many other properties could also be obtained from the same data, such as
 decay branching fractions, mixing angles, and the spin of the sparticle \cite{Berggren:2015qua,Baer:2019gvu,Berggren:2013vfa}.

\section{Conclusions}
The potential for direct discovery of new particles at the ILC could exceed those
of the LHC in certain well-founded scenarios.
This is because the ILC offers a
clean environment without QCD backgrounds, and a  well-defined initial state.
Furthermore, the ILC detectors
will be more precise, will be hermetic, and will not need
to be triggered.
In addition, ILC can be extended in energy and have polarised beams. 

Synergies between ILC and LHC are expected:  the LHC experiments will have higher energy-reach,
while those at ILC
will be more sensitive for subtle signals.
For example, if SUSY is reachable at the ILC, precision measurements can be made.
This input would help interpret any anomalies seen at the LHC, and
might even be what is needed to turn a 3$\sigma$ excess into a discovery of states
  beyond the reach of ILC.



\section{Acknowledgements}
We would like to thank the LCC generator working group and the ILD software
working group for providing the simulation and reconstruction tools and
producing the Monte Carlo samples used in this study.
This work has benefited from computing services provided by the ILC Virtual
Organisation, supported by the national resource providers of the EGI
Federation and the Open Science GRID.



\begin{thebibliography}{99}
  \itemsep0.5ex
\bibitem{Adolphsen:2013kya}
C.~Adolphsen, M.~Barone, B.~Barish, \textit{et al.},
\href{https://arxiv.org/abs/1306.6328}{[arXiv:1306.6328 [physics.acc-ph]]}.

\bibitem{ILC:2007vrf}
T.~Behnke \textit{et al.} [ILC],
\href{https://arxiv.org/abs/0712.2356}{[arXiv:0712.2356 [physics.ins-det]]}.

\bibitem{ILDConceptGroup:2020sfq}
  H.~Abramowicz \textit{et al.} [ILD Concept Group],
  \href{https://arxiv.org/abs/2003.01116}{[arXiv:2003.01116 [physics.ins-det]]}.

\bibitem{Behnke:2013lya}
J.~E.~Brau, \textit{et al.}
``The ILC TDR - Vol. 4, Chap. II : SiD,''
\href{https://arxiv.org/abs/1306.6329}{[arXiv:1306.6329 [physics.ins-det]]}.

\bibitem{Thomson:2009rp}
M.~A.~Thomson,
NIM A \textbf{611} (2009), 25-40
\href{https://arxiv.org/abs/0907.3577}{[arXiv:0907.3577 [physics.ins-det]]}.




\bibitem{Wang:2020lkq}
  Y.~Wang, M.~Berggren and J.~List,
  \href{https://arxiv.org/abs/2005.06265}{arXiv:2005.06265 [hep-ex]}.

\bibitem{Barklow:2017suo}
  T.~Barklow,  \textit{et al.}
Phys. Rev. D \textbf{97} (2018) no.5, 053003
\href{https://arxiv.org/abs/1708.08912}{[arXiv:1708.08912 [hep-ph]]}.


\bibitem{Habermehl:2020njb}
M.~Habermehl,  \textit{et al.}
Phys. Rev. D \textbf{101} (2020) no.7, 075053
\href{https://arxiv.org/abs/2001.03011}{[arXiv:2001.03011 [hep-ex]]}.

\bibitem{Kalinowski:2021tyr}
  J.~Kalinowski, \textit{et al.}
Eur. Phys. J. C \textbf{81} (2021) no.10, 955
\href{https://arxiv.org/abs/2107.11194}{[arXiv:2107.11194 [hep-ph]]}.

\bibitem{Wess:1974tw}
J.~Wess and B.~Zumino,
Nucl. Phys. B \textbf{70} (1974), 39-50.

\bibitem{Roy:2007ay}
D.~P.~Roy,
AIP Conf. Proc. \textbf{939} (2007) no.1, 63-74
\href{https://arxiv.org/abs/0707.1949}{[arXiv:0707.1949 [hep-ph]]}.

\bibitem{Bagnaschi:2017tru}
  E.~Bagnaschi,  \textit{et al.}
Eur. Phys. J. C \textbf{78} (2018) no.3, 256
\href{https://arxiv.org/abs/1710.11091}{[arXiv:1710.11091 [hep-ph]]}.

\bibitem{NunezPardodeVera:2021cdw}
M.~T.~N\'u\~nez Pardo de Vera, M.~Berggren and J.~List,
\href{https://arxiv.org/abs/2105.08616}{[arXiv:2105.08616 [hep-ph]]}.

\bibitem{PardodeVera:2020zlr}
M.~T.~N\'u\~nez Pardo de Vera, M.~Berggren and J.~List,
\href{https://arxiv.org/abs/2002.01239}{[arXiv:2002.01239 [hep-ph]]}.


\bibitem{ATLAS:2018ojr}
M.~Aaboud \textit{et al.} [ATLAS],
Eur. Phys. J. C \textbf{78} (2018) no.12, 995
\href{https://arxiv.org/abs/1803.02762}{[arXiv:1803.02762 [hep-ex]]};

G.~Aad \textit{et al.} [ATLAS],
Phys. Rev. D \textbf{101} (2020) no.5, 052005
\href{https://arxiv.org/abs/1911.12606}{[arXiv:1911.12606 [hep-ex]]};

G.~Aad \textit{et al.} [ATLAS],
\href{https://arxiv.org/abs/2106.01676}{[arXiv:2106.01676 [hep-ex]]};

 [ATLAS],
ATL-PHYS-PUB-2018-048;

\href{http://lepsusy.web.cern.ch/lepsusy/www/inoslowdmsummer02/charginolowdm_pub.html}{LEP~LEPSUSYWG/02-04.1}.
  
\bibitem{Berggren:2015qua}
  M.~Berggren,
 \textit{et al.}
Eur. Phys. J. C \textbf{76} (2016) no.4, 183
\href{https://arxiv.org/abs/1508.04383}{[arXiv:1508.04383 [hep-ph]]}.

\bibitem{Baer:2019gvu}
  H.~Baer, \textit{et al.}
Phys. Rev. D \textbf{101} (2020) no.9, 095026
\href{https://arxiv.org/abs/1912.06643}{[arXiv:1912.06643 [hep-ex]]}.

\bibitem{Berggren:2013vfa}
  H.~Sert, \textit{et al.}
Eur. Phys. J. C \textbf{73} (2013) no.12, 2660
\href{https://arxiv.org/abs/1307.3566}{[arXiv:1307.3566 [hep-ph]]}.






\end{thebibliography}
\end{document}